\documentclass[%
 reprint,
 amsmath,amssymb,
 aps,
]{revtex4-1}

\usepackage{graphicx}
\usepackage{dcolumn}
\usepackage{bm}
\usepackage{here}
\usepackage{color}
\usepackage{soul}
\usepackage[outdir=./]{epstopdf}

\begin{document}

\preprint{APS/123-QED}

\title{Multi-scale microrheology using fluctuating filaments as stealth probes}

\author{Kengo Nishi\textsuperscript{1,2}}
\author{Fred C.\ MacKintosh\textsuperscript{3,4,5}}
\author{Christoph F. Schmidt\textsuperscript{1,2}}

\address{
\textsuperscript{1}Third Institute of Physics-Biophysics, University of G\"{o}ttingen, 37077 G\"{o}ttingen, Germany\\
\textsuperscript{2}Department of Physics \& Soft Matter Center, Duke University, Durham, NC 27708, USA\\
\textsuperscript{3}Department of Chemical \& Biomolecular Engineering, Rice University, Houston, TX 77005, USA\\
\textsuperscript{4}Center for Theoretical Biological Physics, Rice University, Houston, TX 77030, USA\\
\textsuperscript{5}Departments of Chemistry and Physics \& Astronomy, Rice University, Houston, TX 77005, USA
}

\date{\today}

\begin{abstract}
The mechanical properties of soft materials can be probed on small length scales by microrheology. 
A common approach tracks fluctuations of micrometer-sized beads embedded in the medium to be characterized.
This approach yields results that depend on probe size when the medium has structure on comparable length scales. 
Here, we introduce filament-based microrheology (FMR) using high-aspect-ratio semi-flexible filaments as probes. 
Such quasi-1D probes are much less invasive than beads due to their small cross sections. 
Moreover, by imaging transverse bending modes, we simultaneously determine the micromechanical response of the medium on multiple length scales corresponding to the mode wavelengths. 
We use semiflexible single-walled carbon nanotubes (SWNTs) as probes that can be accurately and rapidly imaged based on their stable near-IR fluorescence. 
We find that the viscoelastic properties of sucrose and polymeric hyaluronic acid solutions measured in this way are in good agreement with those measured by conventional micro- and macrorheology. 
\end{abstract}

\pacs{Valid PACS appear here}
\maketitle

Soft materials such as surfactant- or polymer solutions typically have structure at length scales beyond the molecular scale and exhibit characteristic relaxation times from milliseconds to hours or longer. 
These processes can be probed at the macroscopic level by conventional rheology. 
Macroscopic rheology is usually limited, however, to time scales longer than $\sim 0.1$ s \cite{1996_Macosko}.
Various \emph{microrheology} techniques have been developed that can probe in the $\sim 10$ $\mu$s to s range, e.g., using $\mu$m-size particles \cite{1995_Mason_PRL,1997_Gittes_PRL,2000_Crocker_PRL,1999_Schmidt_CurrOpinColloidInterfaceSci,2017_Squires_OxfordUnivPress}. 
Small probes also allow one to study small samples in confined geometries.  Microrheology is sensitive to length scales comparable to or larger than the probe size \cite{2003_Lubensky_PRL,2007_MacKintosh_Science,2014_Weitz_Cell}. 
Simple continuum mechanics might be inappropriate to interpret data, but microrheology can be used to explicitly probe local structure in complex media. 
Correlated fluctuations of pairs of particles can be monitored to probe response on varying length scales (particle distance) \cite{2000_Crocker_PRL,2002_Hough_PRE,2005_Buchanan_PRE}. 
 Artifacts can still be created by the presence of the $\mu$m-sized probes, and, most importantly, particular samples might not be accessible to the probes. This especially holds for biological materials, cells or tissues, where beads are difficult to insert or are actively expelled from e.g. the cell nucleus, the mitotic spindle or the actin cortex  \cite{2003_Yodh_PRL,2004_Valentine_BiophyJ,2011_He_PRE,2014_Atakhorrami_PRL}. 

Here, we introduce the use of slender filaments, in practice semi-flexible polymers, as local \emph{stealth probes}. Filaments embedded in a viscoelastic network, such as microtubules in the cell cytoskeleton  \cite{2007_Brangwynne_PNAS,2008_Brangwynne_PRL}, undulate with the thermal or active motions of the network, but their bending stiffness also affects network fluctuations.
While the filament length is relevant for the hydrodynamic interaction with the embedding medium, the filament diameter determines local perturbations due to excluded volume. 
We use minimally invasive single-walled carbon nanotubes (SWNTs) with extreme aspect ratios: diameter of $\sim 1$ nm and lengths up to tens of $\mu$m (Fig. 1a).
SWNTs have well-defined chemical structures and a precisely known bending stiffness \cite{2001_Yakobson_PRB,2009_Fakhri_PNAS}.
Semiconducting SWNTs exhibit photostable near-IR fluorescence, permitting long-time and high-resolution tracking of their positions and shapes \cite{2009_Fakhri_PNAS}. 
We decompose recorded fluctuating SWNT shapes into dynamic bending eigenmodes (Fig. 1b).  Each mode is sensitive to the medium properties on the scale of its wavelength, similar to the bending modes of a membrane \cite{1975_Brochard_JPhys,2001_Helfer_PRE,2001_Helfer_PRL,2002_Levine_PRE,2011_Granek_SoftMatter}.
By resolving bending modes with wavelengths up to tens of $\mu$m, we simultaneously measure medium response on multiple length scales using a single filament. 
Filament microrheology (FMR) offers advantages over conventional microrheology: (1) Multiple length scales can be probed simultaneously; (2) sensitivity can be tuned by the filament stiffness; and (3) the method can even use endogenous cytoskeletal biopolymers, such as actin filaments and microtubules.

\begin{figure}[htp!]
\centering 
\includegraphics[width=60mm]{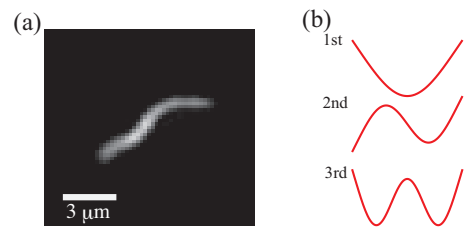}
\caption{(a) A near-infrared fluorescence image of a SWNT in a 4.5 mg/ml hyaluronic acid solution. (b) The first three spatial dynamic eigenmodes of an elastic beam with free ends.}
\end{figure}

We parameterize the shape of a weakly undulating filament at time $t$ by the transverse deflection $u(s,t)$ along its arc length $s$. We describe transverse filament motion by a generalized Langevin equation describing the net force per unit length on the chain at position $s$ \cite{1985_Pecora_Macro,1993_Gittes_JCellBio}:
\begin{equation}
0=-\kappa\frac{\partial^4}{\partial s^4}u(s,t)-\int_{-\infty}^t dt' \alpha(t-t')u(s,t')+\xi(s,t)\,.\label{eqMotion}
\end{equation}
The first term accounts for the elastic restoring force \cite{1993_Gittes_JCellBio}, with bending rigidity $\kappa$. 
The second term is the viscoelastic drag, where the resistance per unit length is given by the memory function $\alpha(t)$, whose
Fourier transform $\alpha(\omega)$ is proportional to the complex shear modulus of the medium, $G(\omega)$ \cite{1986_Landau,2006_Brangwynne_JCB}. 
For the transverse displacement of a rigid rod of length $L$ and diameter $d$ in a viscous liquid, $\alpha(\omega)\simeq-4\pi i\omega \eta/\ln(A L/d)$, where $A\simeq 2.3$, and $\eta$ is the viscosity \cite{1970_Cox_JFluMech,1981_Broersma_JCP,2004_Li_PRE,2007_Brangwynne_BiophysJ}.
The Brownian force  $\xi(s,t)$ has a zero mean $\langle \xi(s,t)\rangle=0$ and a power spectrum satisfying $\langle \xi(s,\omega)\xi(s',\omega)\rangle=\frac{2k_BT}{\omega}\delta(s-s')\rm{Im}[\alpha(\omega)]$, with Boltzmann's constant $k_B$ and temperature $T$, as required by the fluctuation dissipation theorem (FDT).

We expand $u(s,t)$ into orthogonal dynamic eigenmodes $y_q(s)$ as $u(s,t)=\sum_qa_{q}(t)y_{q}(s)$ with wave number $q=\alpha_k/L=\left(k+\frac{1}{2}\right)\pi/L$ for free-end boundary conditions \cite{1985_Pecora_Macro} (see Supporting Information). 
The projection of Eq. \ref{eqMotion} onto a particular spatial mode $y_{q}(s)$ gives the equation of motion for the $k$th mode as $0=-\int_{-\infty}^t dt' \alpha(t-t')a_{q}(t')-\kappa q^4a_{q}(t)+\xi_{q}(t)$. 
Assuming linear response to the Fourier component of the force $f_q$, the amplitude of this mode will be $a_{q}(\omega)=\chi_{q}(\omega)f_{q}(\omega)$, with the response function $\chi_{q}(\omega)=(\kappa q^4+\alpha(\omega))^{-1}$.
The FDT relates the amplitude autocorrelation function $C_{q}(t)=\langle a_{q}(t)a_{q}(0)\rangle$ of each mode $k$ to the corresponding time-dependent response function $\chi_{q}(t)$ for $t>0$: \cite{1987_Chandler,2018_Nishi_SoftMatter}
\begin{equation}
k_BT\chi_{q}(t)=-\frac{d}{dt}\langle a_{q}(t)a_{q}(0)\rangle=\frac{1}{2}\frac{d}{dt}M_{q}(t), \label{eqSym}
\end{equation}
where the mean-squared amplitude difference (MSAD) is defined as $M_{q}(t)=\langle [a_{q}(t)-a_{q}(0)]^2\rangle=\langle[\Delta a_q(t)]^2\rangle$.
Fourier transformation is used to obtain the frequency-dependent response function $\chi_{q}(\omega)=\chi_{q}'(\omega)+i\chi_{q}''(\omega)=\int^{\infty}_0dt\chi_{q}(t)e^{i\omega t}$. 
To increase accuracy, we applied the five-point stencil method to calculate the numerical derivative and Simpson's rule for the subsequent integral \cite{2018_Nishi_SoftMatter}.
The response function $\chi_{q}(\omega)$ is thus calculated from direct integral transforms of the MSAD using the FDT. 
Alternatively, $\chi_{q}(\omega)$ and $G(\omega)$ can be evaluated from the bending fluctuations using a Kramers-Kronig integral (KK integral) \cite{1997_Gittes_PRL}. Detailed and comparisons are given in Fig. S9.
The complex $\alpha(\omega)$ and the complex shear modulus $G(\omega)$ can be evaluated from $\chi_{q}(\omega)$ via: 
\begin{equation}
\chi_{q}(\omega)^{-1}-\kappa q^4=\alpha(\omega)\simeq 4\pi G(\omega)/\ln(AL_{\mbox{\scriptsize eff}}/d),
\end{equation}
where $L{\mbox{\scriptsize eff}}\simeq L/(k+\frac{1}{2})$ is the characteristic length of the undulation \cite{2006_Brangwynne_JCB}. \color{black}
For a SWNT diameter of $\sim$ 1 nm and $L_{\mbox{\scriptsize eff}}$ of $\sim 3$ $\mu$m, 
$\alpha\simeq 1.4\,G(\omega)$.
The real part of Eq. (3) contains the elastic part of the shear modulus; ${\rm{Re}}(\left[\chi_{q}(\omega)\right]^{-1})-\kappa q^4=k_0G'(\omega)$, where $k_0=4\pi/\ln(AL_{\mbox{\scriptsize eff}}/d)$.
It is evident from Eq. (3) that  $G'(\omega)$ is difficult to quantify from $\chi_{q}(\omega)$ if the term $\kappa q^4/k_0$ with $q=(k+1/2)\pi/L$, which strongly depends on mode number $k$, is significantly larger than $G'(\omega)$.
Low modes of long SWNTs are thus suitable for measuring soft materials (see Fig. S3).

\begin{figure}[htp!]
\centering 
\includegraphics[width=60mm]{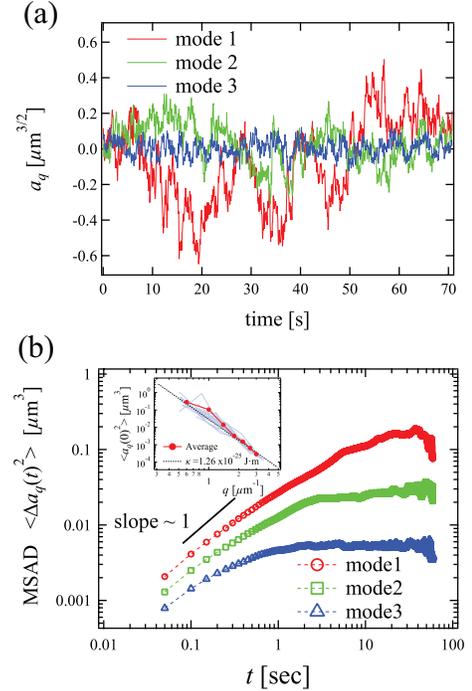}
\caption{(a) Amplitudes of modes 1 to 3 of a 5 $\mu$m SWNT in a 3 mg/ml HA solution. (b) MSADs for the same modes (five recordings averaged). 
Inset: Variance of mode amplitudes plotted vs. wave number (18 recordings from five SWNTs in the same solution, lengths 4.5 $\sim$ 8.5 $\mu$m) \color{black} (blue lines). Averages smoothed by binning (red circles).
Expected dependence $\langle a_{q}(0)^2\rangle=k_BT/\kappa q^4$ with  $\kappa=1.26\times10^{-25}$ J$\cdot$m \cite{2009_Fakhri_PNAS} (black dashed line).}
\end{figure}

We tested our method on two materials: a viscous sucrose solution and a viscoelastic hyaluronic acid (HA) (${\it M}_w=2$ - 2.4 MDa) solution.
Surfactant-wrapped SWNTs were mixed into these solutions and illuminated with a 561 nm laser (see Supporting Information).
SWNT diameter can be determined from its fluorescence spectrum \cite{2002_Bruce_Science}.
Here, a 561 nm laser resonantly excited SWNTs of a (6,5) chirality and 0.78 nm diameter \cite{2005_Tsyboulski_NanoLett, 2009_Fakhri_PNAS}.

Fig. 2(a) shows time series of bending mode amplitudes of a ~ 5 $\mu$m long SWNT in a 3 mg/ml HA solution.
As expected for an intrinsically straight filament, mode amplitudes fluctuated around 0.  The amplitudes decreased for the higher modes as expected from equipartition for thermally excited modes.
Fig. 2(b) shows MSADs for modes 1 to 3.
The bending fluctuations of SWNTs were significantly larger than the noise floor for the first three modes (Fig. S2).
At short times up to 1 s, all MSADs exhibit a power law slope $<$ 1, reflecting the viscoelasticity of the system (Fig. 2b).
At long times, MSADs reach a plateau due to the fact that the filament bending modulus dominates medium response and limits the bending amplitude. MSADs of higher modes reach the plateau earlier due to higher bending energy for a given amplitude. 
From equipartition, the total variance of mode amplitude fluctuations $\langle a_{q}(0)^2\rangle$ should be inversely proportional to the bending rigidity and the 4th power of wave number \cite{1993_Gittes_JCellBio,2007_Brangwynne_BiophysJ}: $\langle a_{q}(0)^2\rangle=k_BT/\kappa q^4$.
This prediction is also plotted in the inset of Fig. 2(b), matching our data using $\kappa=1.26\times10^{-25}$ J$\cdot$m from Ref. \cite{2009_Fakhri_PNAS} (this value is used for further analysis).

To quantitatively confirm FMR, we applied it to a 60 wt\% sucrose solution, a Newtonian fluid. 
We analyzed 24 movies of 8 fluctuating SWNTs with lengths of 4.5 $\sim$ 6 $\mu$m in this solution.
$G''(\omega)$ obtained from the first two bending modes is shown in Fig. S5. 
The values of $G''(\omega)$ from both modes collapse onto a single curve with a power-law slope of $\sim$1 as expected for a Newtonian fluid.
We also measured $G''(\omega)$ of this solution by macroscopic rheology with parallel-plate geometry and found good agreement (Fig. S5).

To confirm FMR in a viscoelastic material, we studied HA solutions with concentrations from 1 to 4.5 mg/ml. HA is an anionic glycosaminoglycan with non-trivial viscoelasticity prevalent in the pericellular matrix of cells \cite{2008_Nijenhuis_Biomacro}.
SWNT fluctuations and tracking results in a 4.5 mg/ml HA solution are shown in Fig. S6.
Thermal fluctuations of the filaments were clearly visible and 
complex shear moduli could be calculated (Fig. 3(a) and Fig. S7).
Again, the higher modes of shorter SWNTs are not suitable to measure $G'(\omega)$ because the term $\kappa q^4/k_0$ becomes dominant over $G'(\omega)$.
Since the low-frequency $G'(\omega)$ of the HA solutions is $\sim$ 0.3 Pa, we chose mode numbers that fulfill $\kappa q^4/k_0\lesssim0.3$ Pa with $q=(k+1/2)\pi/L$ to estimate $G'(\omega)$. 
Complex shear moduli calculated from several modes agree at each concentration as expected since the response of HA solutions does not depend on length scale in the $\mu$m range.
Results from filament bending dynamics ($G_{\rm{filament}}$) were compared with those obtained from bead microrheology ($G_{\rm{bead}}$), showing good agreement (Fig. 3(a) and Fig. S7)
(see Supporting Information).
In the high-frequency regime, $G_{\rm{bead}}$ extends beyond $G_{\rm{filament}}$ by as much as half a decade in frequency because of frame rate differences, 50 Hz for bead microrheology and 10/20 Hz for FMR.
For an exact comparison between $G_{\rm{bead}}$ and $G_{\rm{filament}}$, real and imaginary parts of $G_{\rm{bead}}$ were fitted by power-law functions, and both components of $G_{\rm{filament}}$ derived from multiple bending modes were normalized by  by the bead results (Fig. S8). Results are close to 1, indicating that $G_{\rm{filament}}$ gives results consistent with bead microrheology.

Note that thermal bends of a filament in a polymer network can relax either by following the surrounding network or by reptation, i.e. anisotropic diffusion through the network \cite{1967_Edwards,1971_deGennes,1978_Doi,1994_Kas_Nature,1996_Kas_BiophysJ,2010_Fakhri_Science}. 
Reptation will contribute to mode-amplitude relaxation if it occurs rapidly enough to compete with network dynamics \cite{1978_Doi,1998_Morse_Macro}. We neglect reptation here since in the viscous or weakly elastic solutions we probed, medium relaxation was dominant. In more strongly entangled polymer networks, reptation needs to be taken into account or suppressed by crosslinking the probe filament to the network.
Shorter-wavelength modes should be less affected by reptation artefacts than longer-wavelength modes. 
In our data, the viscoelasticity evaluated from different modes coincides, and the results also coincide with those from conventional micro/macrorheology, proving that 
reptation was negligible in our experimental time window, and that the transverse filament bending modes accurately reported medium viscoelasticity. 

\begin{figure}[htp!]
\centering 
\includegraphics[width=80mm]{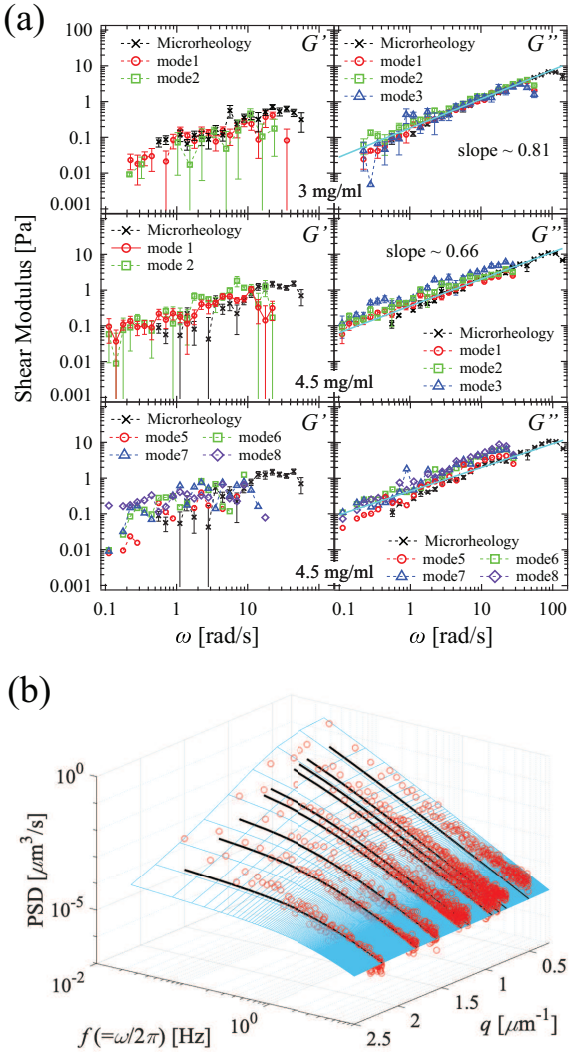}
\caption{(a) Viscoelasticity of HA solutions measured from the bending dynamics of SWNTs. 
HA concentrations given in graphs. 
Upper row: 18 recordings from 5 SWNTs, lengths 4.5 $\sim$ 8.5 $\mu$m. Middle row: 14 recordings from 8 SWNTs, lengths 4.5 $\sim$ 7.5 $\mu$m. Bottom row: 1 SWNT, length 19 $\mu$m. Shear elastic moduli of each HA solution were also measured by conventional bead microrheology (black crosses). 
Power-law fits of $G''_{\rm{bead}}$ microrheology are shown as solid light blue lines.
(b) Global 2D fit of PSDs as a function of wave number $q$ and frequency $f (=\omega/2\pi)$ (10 recordings from three SWNTs with lengths of 9.15, 6.13, and 5.03 $\mu$m). Three modes (1 to 3) of each SWNT are plotted. The fit with Eq. (4) is shown as light-blue mesh plane. Black solid lines represent slices of the fitted plane at $q = 0.515$, 0.769, 0.858, 0.937, 1.20, 1.28, 1.56, 1.79, and 2.19 $\mu $m$^{-1}$, which correspond to the wave numbers of three modes (1 to 3) of SWNTs with lengths of 9.15, 6.13, and 5.03 $\mu$m, respectively.}
\end{figure}

Complex shear moduli increased with increasing HA concentration (Fig. 3).
To monitor changes in the frequency dependence of rheological properties, we fitted $G''_{\rm{bead}}$ with power laws (Fig. 3(a)), finding a slope close to 1 for 1 mg/ml HA is, with the slope decreasing with increasing HA concentration, consistent with previous studies \cite{2008_Nijenhuis_Biomacro}.
Above the entanglement concentration $c_e (\gg c^*)$, an elastic plateau region is expected \cite{1988_Doi}, where $c^*$ is the overlap concentration, estimated to be $\sim 0.3$ mg/ml.
We did not observe a plateau, even in the 4.5 mg/ml HA solution, implying weak entanglement.

Fig. 3(a) shows data obtained from a 19 $\mu$m long SWNT in the 4.5 mg/ml HA solution, where we could evaluate 8 modes.
For modes 1-4, the recording time was too short compared to mode-relaxation times so that they could not be used (see Supporting Information and Fig. S10). $G_{\rm filament}$ from modes 5-8 are consistent with $G_{\rm{bead}}$.
Note that the effective length of mode 8 ($L_{\mbox{\scriptsize eff}} = 2.2$ $\mu\rm{m}$) is approximately 5.7 times shorter than that of mode 1 ($L_{\mbox{\scriptsize eff}} = 12.7$ $\mu\rm{m}$), illustrating the possibility to measure medium response over a  range of length scales with a single filament.

Eq. (3) shows that the mode dynamics depend both on the response of the embedding medium and the filament bending stiffness.  When filament properties are known, our approach thus allows us to measure medium response. We can alternatively obtain filament stiffness if medium response is known. As a consistency check, we performed a global 2D-fit on nine different power spectral densities (PSD) of mode amplitudes $\langle |a_{q}(\omega)|^2\rangle$.  The PSD is the Fourier transform of the MSAD. We used modes 1,2,3 of three SWNTs with lengths of 9.15, 6.13, and 5.03 $\mu$m in 3 mg/ml HA to obtain SWNT bending stiffness (Fig. 3(b)), using the medium response $G_{\rm bead}$ measured by bead microrheology. Towards low frequencies, PSDs level off for the higher $q$ modes when the filament bending modulus increasingly restricts the thermal bending amplitudes.
The scale-dependent PSDs plotted in Fig. 3(b) can be described by the generalized Langevin equation, Eq. (1). Switching to Fourier space and using for the power spectrum of the Brownian force, $\langle \xi(s,\omega)\xi(s',\omega)\rangle=\frac{2k_BT}{\omega}\delta(s-s')\rm{Im}[\alpha(\omega)]$, gives the PSDs as: $\langle |a_{q}(\omega)|^2\rangle=\frac{k_BT}{\omega}{\rm{Im}}[\chi_{q}(\omega)]=\frac{k_BT}{\omega}\frac{\alpha''(\omega)}{|\kappa q^4+\alpha(\omega)|^2}$ (Eq. (4)) with $q=(k+1/2)\pi/L$ for mode number $k$.
We can thus globally fit all the scale-dependent PSDs with a 2D plane defined by Eq. (4) (Fig. 3(b)), with just one free parameter, $\kappa$, having fixed the memory function $\alpha(\omega)=k_0G(\omega)$ with the power-law fitting results of $G_{\rm bead}$.
Slices of the fitted plane can be compared with the data.
From this fit, we found $\kappa=(7.09\pm1.04)\times10^{-26}$ J$\cdot$m, close to the reported value $\kappa=1.26\times10^{-25}$ J$\cdot$m \cite{2009_Fakhri_PNAS}.

To quantify the stealth character of filament microrheology, we can estimate how local depletion and non-affine deformations around the probe \cite{2003_Yodh_PRL,2004_Valentine_BiophyJ,2011_He_PRE,2014_Atakhorrami_PRL}, would affect results for shear moduli. Probe geometry enters through Eq. (3) in the relation $\alpha(\omega)=k_0G(\omega)$ with, in the case of a filament as probe, $k_0\approx4\pi/\ln(AL_{\mbox{\scriptsize eff}}/d)$.  For a SWNT with 0.78 nm diameter and 10 $\mu$m length , the error of the shape factor $k_0$ would be less than 8$\%$ even if the effective diameter of the filament were to double due to local non-affine deformation effects. Because $k_0$ is inversely proportional to the logarithm of its aspect ratio, FMR is thus quite insensitive to local perturbations due to filament cross-section.

In conclusion, we have introduced and tested filament microrheology (FMR) as a new method to measure shear elastic moduli in soft viscoelastic media, evaluating the bending dynamics of embedded filaments.
Slender filaments with two dimensions on the nm scale and lengths on the $\mu$m scale cause minimal local perturbations, easily penetrate dense media such as the cell-internal structures or the nucleus, while still coupling to mesoscopic medium dynamics on the $\mu$m scale. 
Furthermore, filaments report complex shear moduli at multiple length scales simultaneously.
FMR is thus uniquely useful to measure scale-dependent viscoelasticity of soft materials with hierarchical structures, for example the cytoskeleton of living cells.
In our samples we found good agreement with conventional microrheology over almost two orders of magnitude in frequency. 
We expect that other semi-flexible filaments such as actin filaments or microtubules can be used as probe filaments in biological systems, 
which would be entirely non-invasive and completely avoid the introduction of foreign objects into cells.
Our approach also suggests possible extensions using the shape fluctuations of other extended objects such as membranes \cite{1975_Brochard_JPhys,2001_Helfer_PRE,2001_Helfer_PRL,2002_Levine_PRE,2011_Granek_SoftMatter} to quantify the rheological properties of the surrounding medium.
\\
\acknowledgements 
This research was supported in part by the European Research Council under the European Union's Seventh Framework Programme (FP7/2007- 2013) / ERC grant agreement no. 340528 (to CFS). FCM was supported in part by the National Science Foundation Division of Materials Research (Grant No. DMR-1826623) and the National Science Foundation Center for Theoretical Biological Physics (Grant No. PHY-2019745). 

\bibliography{filamentMRver1}

\clearpage

\begin{center}
\textbf{\large Supplement: Multi-scale microrheology using fluctuating filaments as stealth probes}
\end{center}

\renewcommand{\thefigure}{S\arabic{figure}}
\setcounter{figure}{0}

\section{Materials and Methods}

Sodium hyaluronate with a weight-average molecular weight (${\it M}_w$) of 2 - 2.4 MDa, sucrose, and sodium deoxycholate (NaDOC) were obtained from Sigma Aldrich Corp. (St. Louis, MO, USA). Phosphate-buffered saline (PBS) was obtained from Invitrogen (Carlsbad, CA, USA). SWNTs produced in a HiPco reactor were obtained under a MTA from Rice University (batch number 189.2).

About 1 mg SWNTs was mixed with 2 mL of 2 wt \% NaDOC solution in a glass scintillation vial. The vial was sonicated (Vibra Cell, VC-50; Sonics and Materials, Newtown, CT, USA) at a power of 5 W for 7 - 8 s using a 2-mm diameter microprobe tip. After sonication, the sample was centrifuged at 300 g for 15 min. The supernatant was carefully collected and stored as a stock solution.

A sucrose solution was chosen as a viscous control. The stock solution of SWNTs was 100x diluted with deionized water. 1 $\mu$L of the SWNT solution was added to 10 $\mu$L of a sucrose solution to prepare a 60 wt\% sucrose solution.

Various concentrations of sodium hyaluronate (HA) dissolved in PBS were prepared as viscoelastic samples. 
First, 5 mg sodium hyaluronate were dissolved in 1 ml PBS and maintained as a stock solution at 4 ${}^{\circ}$C. 
This stock solution was consumed within 2 days to minimize hydrolysis. 
The stock solution of SWNTs was diluted 100x with PBS. 
Samples at different HA concentrations were prepared by diluting the HA stock solution with PBS and adding the dilute SWNT solution. 
The final dilution of SWNT in the samples was more than 1000x from the initial stock SWNT suspension.

The samples were sandwiched between two coverslips using strips of double-stick tape. The chambers were sealed using VALAP (1:1:1, vaseline:lanolin:paraffin). All experiments were performed at room temperature (23 ${}^{\circ}$C).

NIR fluorescence images of individual nanotubes were recorded under a Zeiss Examiner.Z1 upright microscope or a custom-built inverted microscope, both of which equipped with a high-NA objective (alpha Plan-Apochromat, 100x, NA = 1.46; Zeiss). For both microscopes, fluorescence excitation was done with a 561 nm DPSS laser (500 mW cw; Cobolt JiveTM; Cobolt) that was circularly polarized using a quarter-wave plate (AQWP05M-600; Thorlabs) and then focused into the back aperture of the objective. In both microscopes, NIR fluorescence images were taken by a NIR camera with an InGaAs detector (X-Cheetah1 10-CL-TE3, Xenics). A tube lens (f$_T$ = 164.5 mm; Zeiss) focused the light onto the NIR camera in the custom-built inverted microscope. For the Zeiss Examiner.Z1, the NIR camera was connected to a camera side port. The frame rates for video recordings from the sucrose sample and the HA samples were 10 Hz and 20 Hz, respectively. 

Macrorheology was performed on the sucrose solution by a commercial rheometer (MCR501, Anton Paar, Austria) in an oscillatory mode with 5 $\%$ strain at 25 ${}^{\circ}$C using parallel-plate geometry (samples were prepared in parallel with the ones used for filament microrheology). The viscoelasticity of the HA solution was examined by conventional video-microrheology as control, using micron-sized beads.  We used red fluorescent beads (Fluoro-Max R0100, Thermo Fisher Scientific) with a diameter of 0.6 or 1 $\mu$m and recorded with a frame rate of 50 Hz using a high-speed CMOS camera (SA1.1, Photron, Bucks, UK) in the two microscopes described above. We determined positions and trajectories of the fluorescent beads using the Mosaic plugin of ImageJ \cite{2005_Koumoutsakos_JStruBio}.

We recorded and analyzed 24 movies of 8 fluctuating nanotubes with lengths of $4.5\sim6$ $\mu$m in the sucrose solution, 14 movies from 6 nanotubes with lengths of $3.5 \sim 10$ $\mu$m in 1 mg/ml HA , 43 movies from 9 nanotubes with lengths of $6.5 \sim 11$ $\mu$m in  2 mg/ml HA, 18 movies from 5 nanotubes with lengths of $4.5 \sim 8.5$ $\mu$m in 3 mg/ml HA, and 14 movies from 8 nanotubes with lengths of $4.5 \sim 7.5$ $\mu$m in 4.5 mg/ml HA.

\section{Image analysis}

The coordinates of the backbone of CNTs were determined from each image by the JFilament plugin in ImageJ \cite{2010_Vavylonis_Cytoskeleton} and used for further data processing. 
After extracting the backbone coordinates $(x_i,y_i)$ of SWNTs from each images, the amplitude $a_q(t)$ of the $k$th mode was calculated from the local tangent angle of the nanotube \cite{1993_Gittes_JCellBio}.
Here, the amplitude $a_q(t)$ was estimated from the local tangent angle $\theta(s_i)=\tan^{-1}(y_{i+1}-y_i)/(x_{i+1}-x_i)$ via an integration by parts $a_q(t)=\int^{L}_{0}ds{\hspace {0.2pc}}u(s,t)y_q(s)=-\int^{L}_{0}ds{\hspace {0.2pc}}\theta(s,t)\tilde{y}_q(s)$ based on the relation $\partial u(s,t)/\partial s\approx\theta(s,t)$.
 $\tilde{y}_q(s)$ denotes the integral of $y_q(s)$.
The eigenmodes for Eq. (1) are given by $y_q(s)$ as 

\[
y_{q}(s)=\frac{1}{\sqrt[]{\mathstrut L}} \begin{cases}
\frac{\cosh\bigl(q\left(s-\frac{L}{2}\right)\bigr)}{\cosh\bigl(\frac{\alpha_k}{2}\bigr)}+\frac{\cos\bigl(q\left(s-\frac{L}{2}\right)\bigr)}{\cos\bigl(\frac{\alpha_k}{2}\bigr)} & \mbox{($k$ odd)}\\ %
\frac{\sinh\bigl(q\left(s-\frac{L}{2}\right)\bigr)}{\sinh\bigl(\frac{\alpha_k}{2}\bigr)}+\frac{\sin\bigl(q\left(s-\frac{L}{2}\right)\bigr)}{\sin\bigl(\frac{\alpha_k}{2}\bigr)} & \mbox{($k$ even) .}
\end{cases}
\]

For free-end boundary condition, the integrals of the eigenfunctions are given by \cite{1985_Pecora_Macro}

\[
\tilde{y}_q(s)=\frac{\sqrt[]{\mathstrut L}}{\alpha_k} \begin{cases}
\frac{\sinh\bigl(q\left(s-\frac{L}{2}\right)\bigr)}{\cosh\bigl(\frac{\alpha_k}{2}\bigr)}+\frac{\sin\bigl(q\left(s-\frac{L}{2}\right)\bigr)}{\cos\bigl(\frac{\alpha_k}{2}\bigr)} & {\rm if}{\hspace {0.5pc}}k{\hspace {0.5pc}}{\rm odd}\\ %
\frac{\cosh\bigl(q\left(s-\frac{L}{2}\right)\bigr)}{\sinh\bigl(\frac{\alpha_k}{2}\bigr)}-\frac{\cos\bigl(q\left(s-\frac{L}{2}\right)\bigr)}{\sin\bigl(\frac{\alpha_k}{2}\bigr)} & {\rm if}{\hspace {0.5pc}}k{\hspace {0.5pc}}{\rm even}
\end{cases}\,.
\]

During video recording, parts of a given SWNT can fluctuate out of the focal plane due to 3D motions. 
This typically happens at the ends of a filament and causes an apparent length decrease (Fig. S1). 
When the ends of a filament fluctuate in and out of the focal plane, the decomposition into eigenmodes would thus deliver eigenmodes of varying wavevectors. 
The time course of the apparent length changes of an exemplary SWNT is shown in Fig. S1. 
The length was calculated in each frame by tracking the backbone of the SWNT. 
In this particular time course, we observed a strong apparent shortening between frame 1250 and frame 1450. 
We therefore cut  this recording into three intervals, the boundaries of which are marked as dotted vertical lines in Fig. S1. 
Period 2, when one end of the SWNT went out of focus due to out-of-plane bending, was cut out from the data and the two remaining fragments were evaluated as independent shorter recordings.

\begin{figure}[H]
\centering 
\includegraphics[width=70mm]{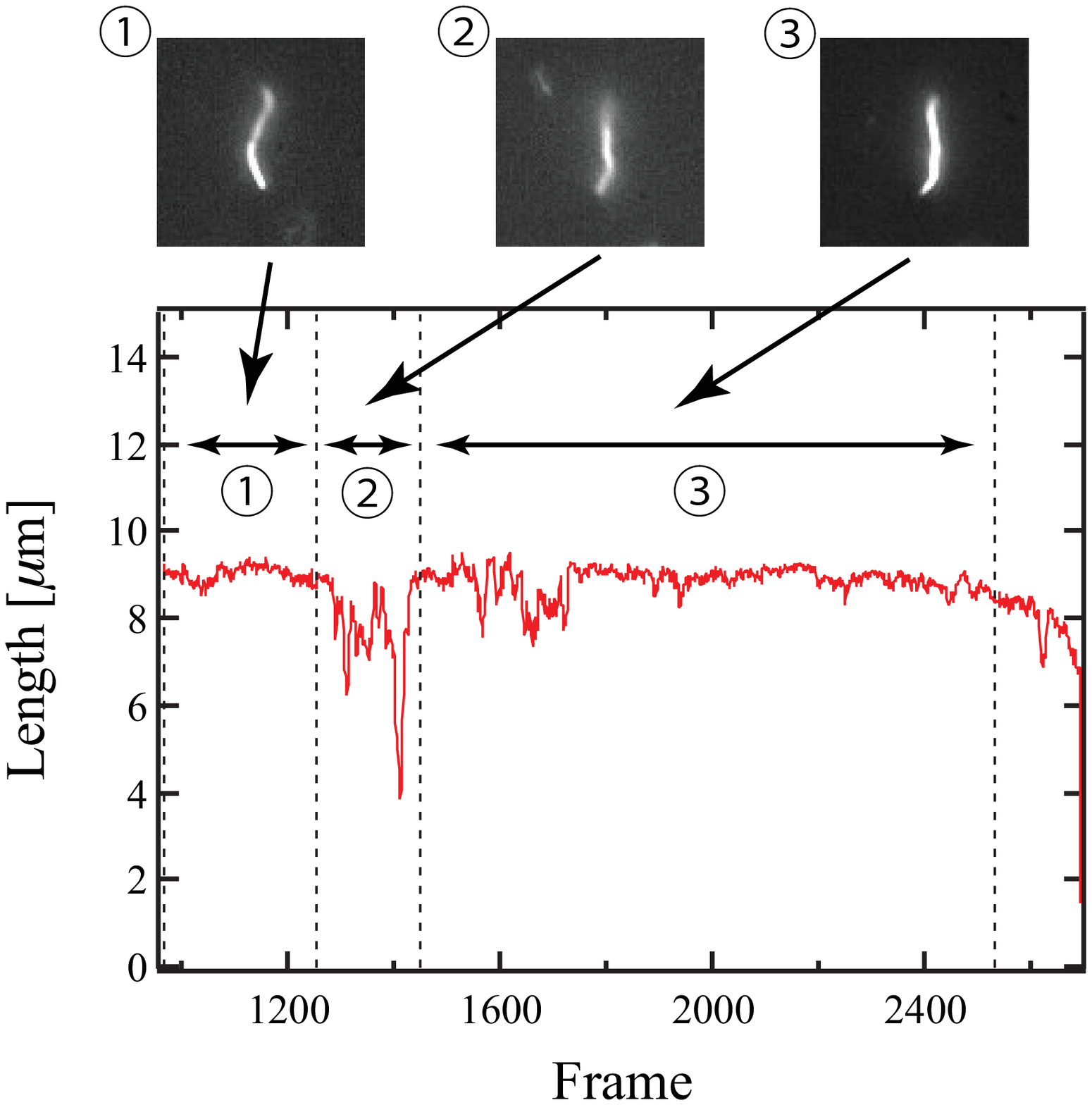}
\caption{Apparent length changes of a filament due to fluctuations out of the focal plane.}
\end{figure}

\section{The effects of noise}

\begin{figure}[H]
\centering 
\includegraphics[width=70mm]{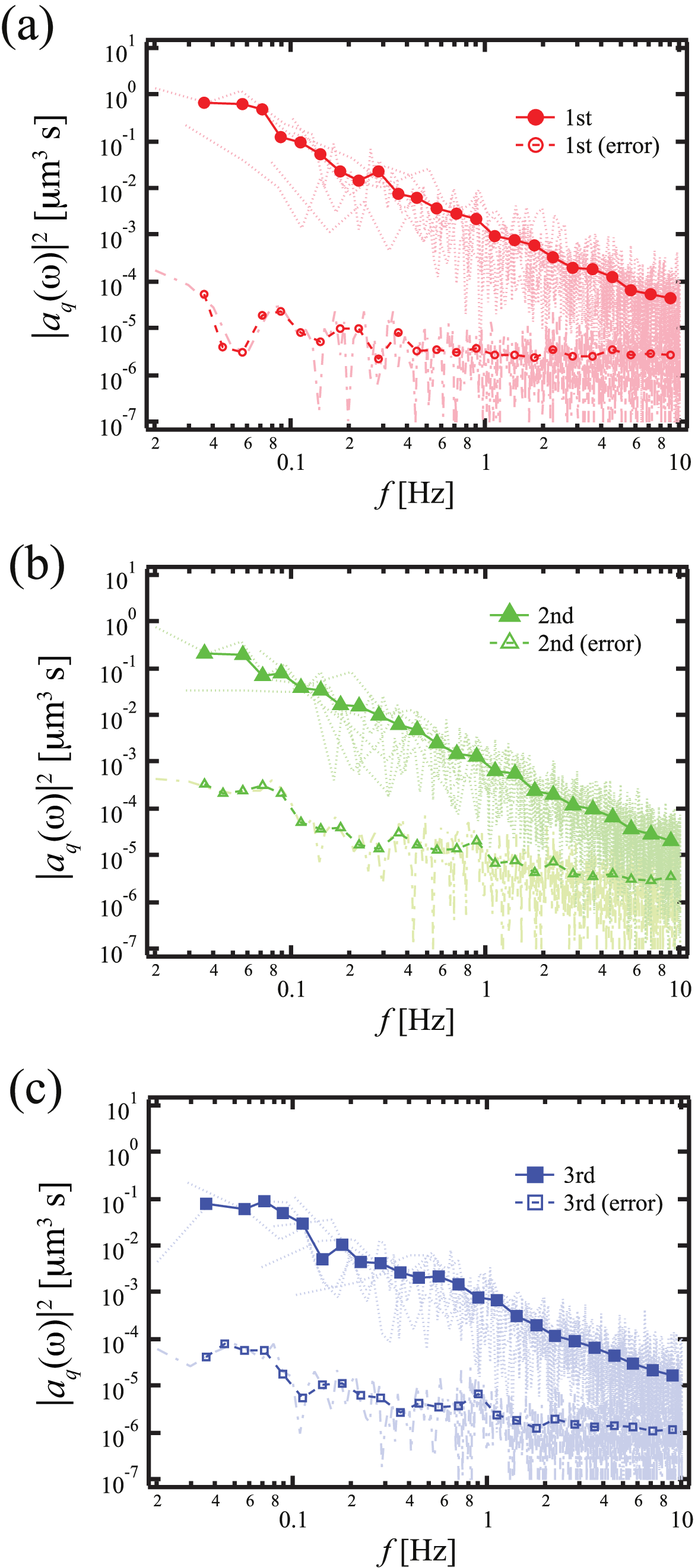}
\caption{Mode amplitude fluctuations of a freely diffusing SWNT compared to detection and tracking noise estimated from surface-immobilized SWNT of equal length. PSDs of (a) amplitude fluctuations of mode 1, (b) mode 2, and (c) mode 3. For this plot, we recorded 6 movies of a 8 $\mu$m SWNT and calculated PSDs of the three lowest eigenmodes, depicted as dotted lines (1st: red, 2nd: green, 3rd: blue). Averages of the 6 movies for each mode were calculated and smoothed by logarithmic binning and are plotted as well (1st: red filled circles, 2nd: green filled triangles, 3rd: blue filled squares).  For comparison: PSDs of the amplitude fluctuations of the first three eigenmodes of a surface-immobilized 8 $\mu$m SWNT on the surface of a cover slip. Data evaluated as above (dashed lines and open symbols). We tracked two movies that consist of 2000 frames.}
\end{figure}

To estimate the noise generated by image pixelation, image resolution limits as well as digitization and tracking, we performed the mode analysis on SWNTs physically attached to the glass coverslip surface. 
We plotted power spectral densities (PSDs) of apparent fluctuation amplitudes of an immobilized SWNT in Fig. S2. 
PSDs of freely fluctuating SWNTs are significantly larger than those of an immobilized SWNT up to the Nyquist frequency of 10 Hz for the first three modes.

\section{Relative importance of filament elasticity and medium elastic response}
\begin{figure}[H]
\centering 
\includegraphics[width=70mm]{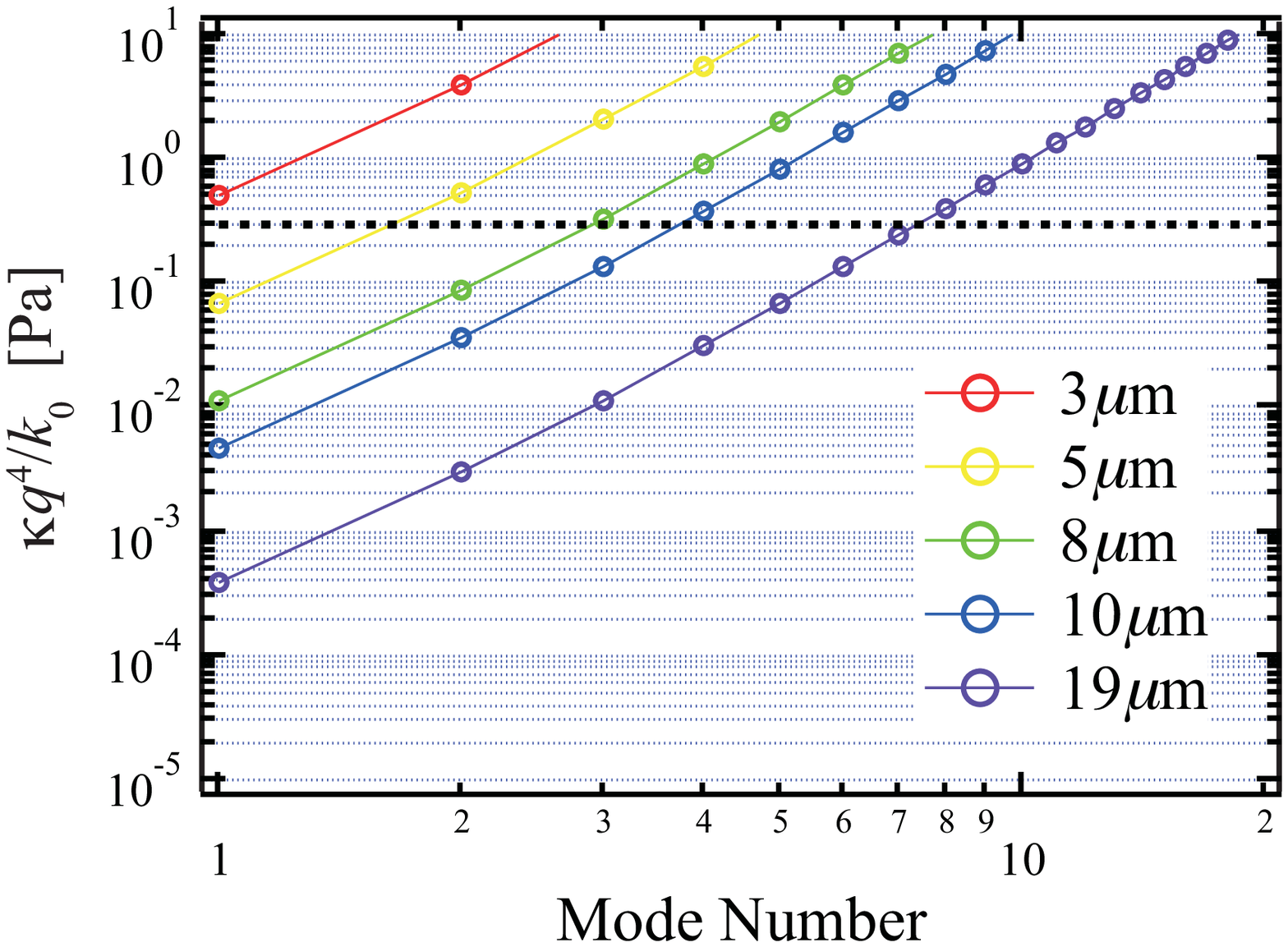}
\caption{Calculated dependence of the term $\kappa q^4/k_0$ on SWNT length and mode number}
\end{figure}

Eq. (3) shows how both medium elasticity and filament bending stiffness affect the filament response. To compare the two contributions, values of the term $\kappa q^4/k_0$ were calculated for eigenmodes up to the 20th mode and for different filament lengths, assuming a filament bending stiffness of $\kappa=1.26\times10^{-25}$ J$\cdot$m
We plotted this apparent $G'(\omega)$ caused by the bending stiffness of SWNTs against mode number for filaments of various lengths using $\kappa=1.26\times10^{-25}$ J$\cdot$m \cite{2009_Fakhri_PNAS} (Fig. S3). 
With increasing mode number and shorter filament lengths, the term $\kappa q^4/k_0$ increases, indicating that the lower modes of longer SWNTs are most suitable for measuring softer materials. 
To pick the appropriate mode numbers with $\kappa q^4/k_0\lesssim0.3$ Pa, a horizontal dashed line of 0.3 Pa is plotted in Fig. S3, and we chose mode numbers for given filament lengths that keep the bending term below that line.

\section{{Variance of bending mode amplitudes}}
\begin{figure}[H]
\centering 
\includegraphics[width=90mm]{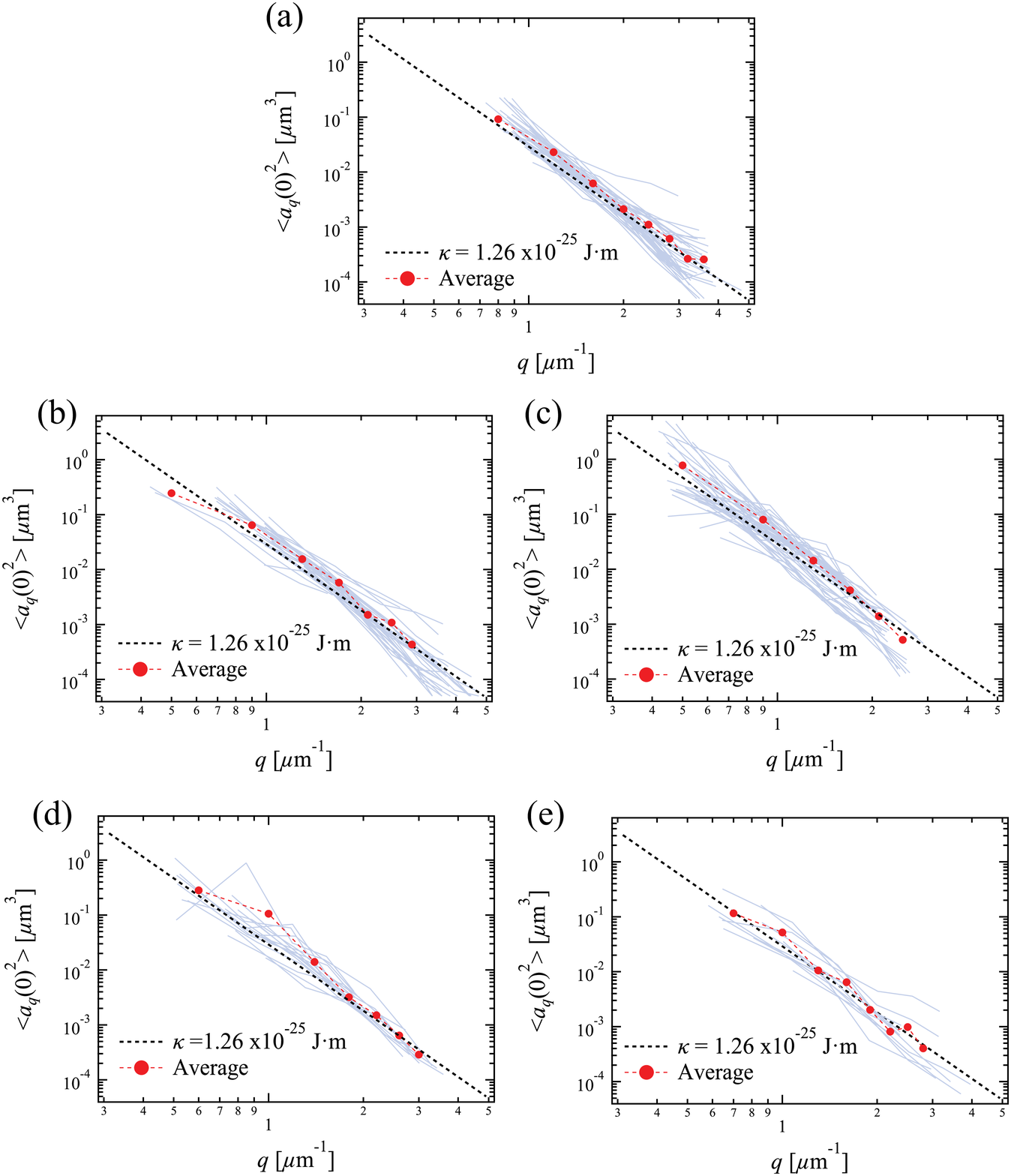}
\caption{Amplitude variance of the bending modes, plotted vs wave number for (a) sucrose solution, (b)1 mg/ml HA, (c) 2 mg/ml HA, (d) 3 mg/ml HA, (e) 4.5 mg/ml HA.}
\label{fig_mode}
\end{figure}

The total variance of amplitude fluctuations for sucrose and HA solutions is plotted against wavenumber in Fig \ref{fig_mode}.
As mentioned above, we recorded and analyzed more than 10 movies of SWNTs in each solution, depicted as blue lines.
Averages of all the movies in each panel were calculated and smoothed by binning and are plotted as red circles.
We also plotted the prediction of the equipartition theorem: $\langle a_q(0)^2\rangle=k_BT/(\kappa q^4)$ with $\kappa=1.26\times10^{-25}$ J$\cdot$m reported in Ref. \cite{2009_Fakhri_PNAS}.
Good agreement was obtained for all the solutions, suggesting that the equipartition theorem is valid in this wavenumber regime.

\section{Microrheology analysis}

We performed bead-based microrheology to obtain reference values for the shear moduli of the media we used $G(\omega)$. 
As discussed in  \cite{2018_Nishi_SoftMatter},  the time-dependent response function $\chi(t)$ can be evaluated from a direct transform of the MSD ($M(t))$ of the thermally fluctuating particles, which is analogous to Eq. 2.
\[
k_BT\chi(t)=\frac{1}{2}\frac{d}{dt}M(t)\,
\]
The frequency-dependent response function $\chi(\omega)$ is then obtained by 
\[
\chi(\omega)=\int^{\infty}_0dt\chi(t)e^{i\omega t}=\chi'(\omega)+i\chi''(\omega) \,.
\]
The complex shear modulus $G(\omega)$ was calculated by both methods via the generalized Stokes formula $G(\omega)=\left(6\pi R \chi(\omega)\right)^{-1}$, where $R$ is the radius of the beads.

\section{Viscosity of a sucrose solution from FMR}
\begin{figure}[htp!]
\centering 
\includegraphics[width=90mm]{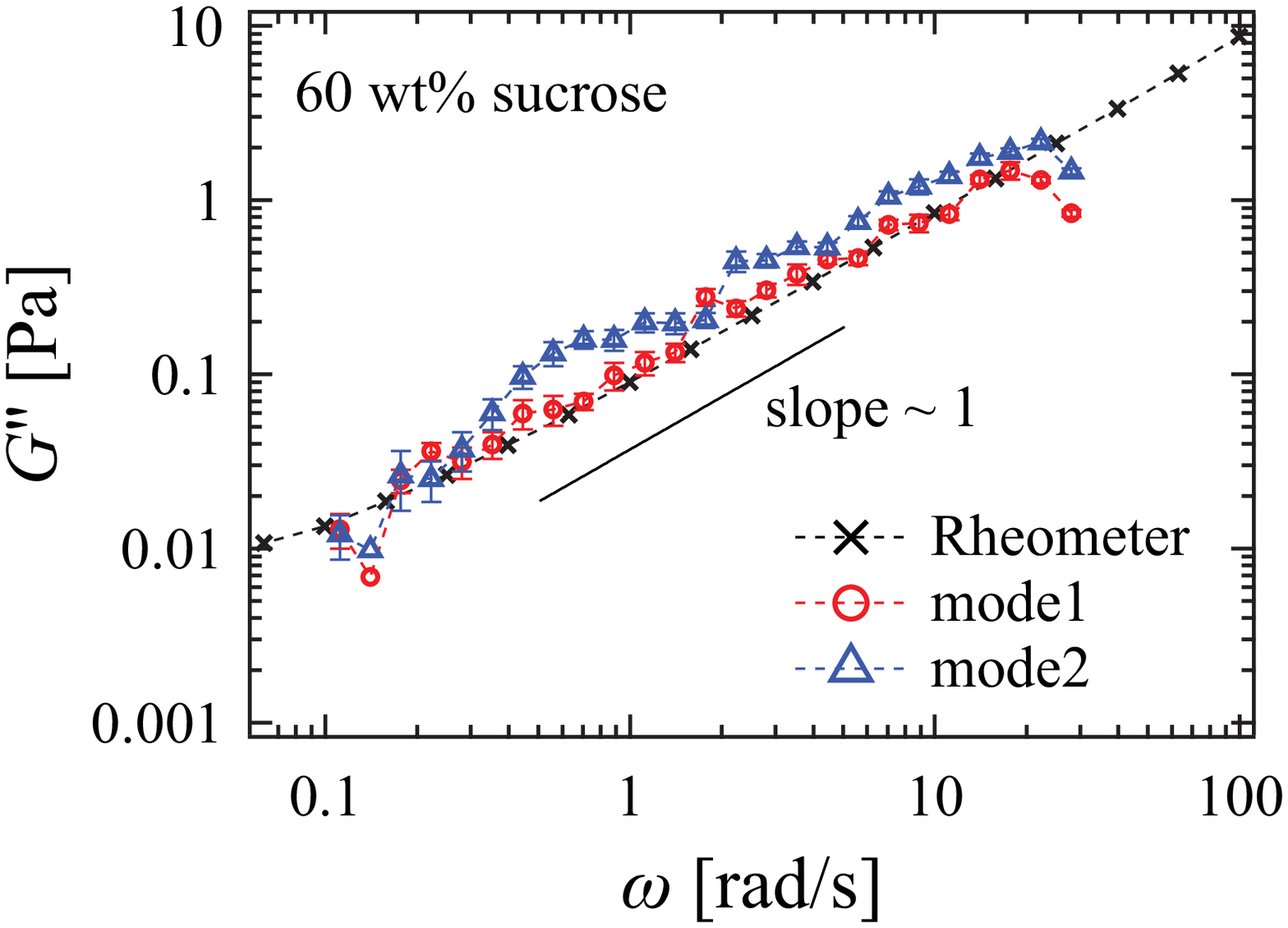}
\caption{Viscosity of a 60 wt\% sucrose solution measured from the bending dynamics of SWNTs, using the lowest two bending. For this plot, we recorded and took averages of 24 movies of 8 fluctuating nanotubes with the lengths of 4.5 $\sim$ 6 $\mu$m.  For comparison, parallel-plate macrorheometry data from the same solution (black crosses). }
\end{figure}

Fig. S5 shows  $G''(\omega)$ of a 60 wt$\%$ sucrose solution measured by FMR and macrorheology. 
The complex shear moduli calculated from the lowest two modes of the fluctuating SWNTs coincide with the macrorheology result. 
The slope of $G''(\omega)$ is close to 1, as expected for a Newtonian liquid.

\section{Single-filament dynamics in a 4.5 mg/ml HA solution}
\begin{figure}[H]
\centering 
\includegraphics[width=60mm]{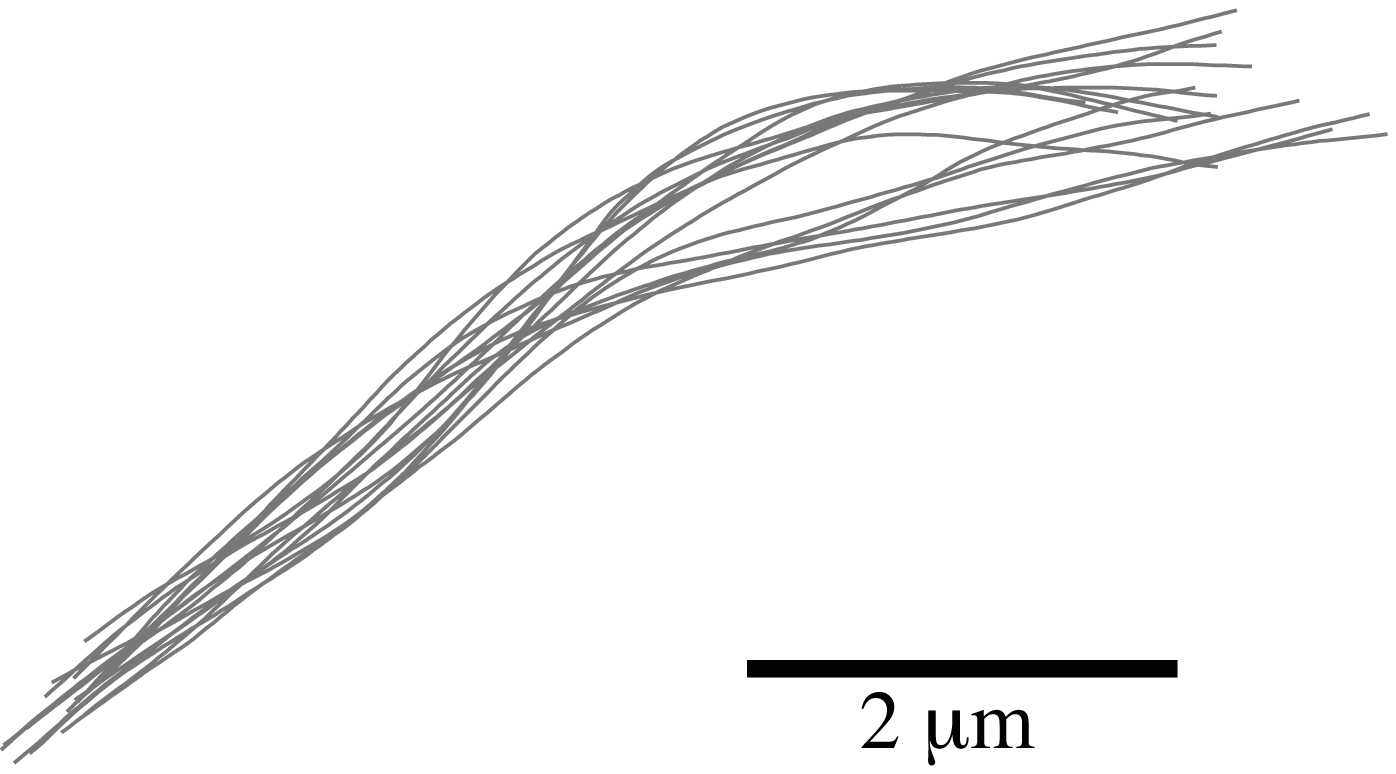}
\caption{Snapshots (15) of contours of a single SWNT performing Brownian motion in a 4.5 mg/ml HA solution, taken at time intervals of 4 s.}
\label{fig_track}
\end{figure}

Tracking results of a SWNT thermally fluctuating in a 4.5 mg/ml HA solution are shown in Fig. \ref{fig_track}. Bending fluctuations of the filament are clearly visible.

\section{Complex shear moduli of 1 and 2 mg/ml HA solutions}

\begin{figure}[H]
\centering 
\includegraphics[width=90mm]{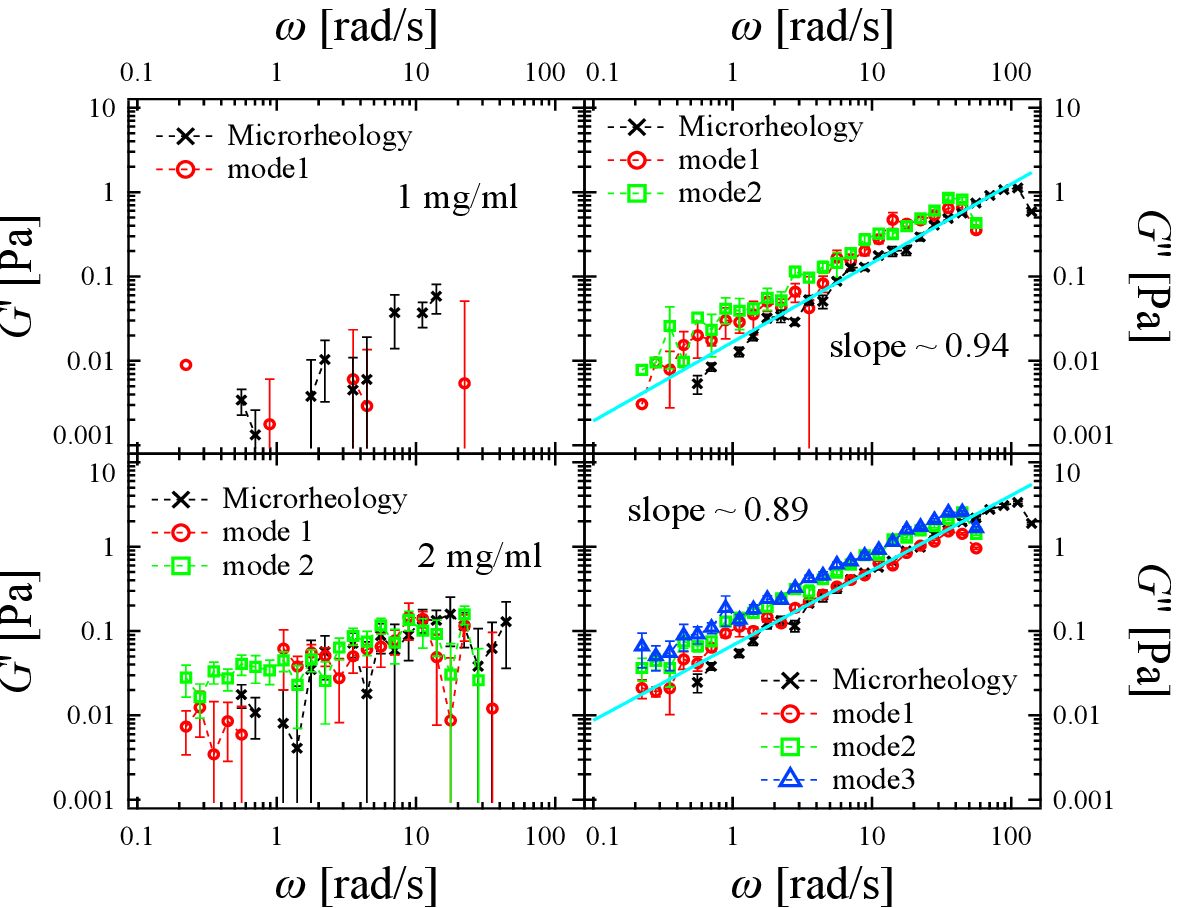}
\caption{Viscoelasticity of 1 and 2 mg/ml HA solutions measured by FMR, using SWNTs (14 recordings from 6 SWNTs with lengths of 3.5 $\sim$ 10 $\mu$m in 1 mg/ml and 43 recordings from 9 SWNTs with lengths of 6.5 $\sim$ 11 $\mu$m in 2 mg/ml) . }
\label{fig_HAlow}
\end{figure}

In addition to the data shown in Fig. 3 in the main text, we plot the complex shear moduli of 1 and 2 mg/ml HA solutions measured by FMR from the bending dynamics of SWNTs and by bead MR (Fig. \ref{fig_HAlow}).
 Complex shear moduli from different modes coincide with the bead-based microrheology results. Power-law fits to the data suggest that the slope of $G''$ increases as the polymer concentration decreases. For the 1 mg/ml HA solution, the slope is close to 1, which is expected for a Newtonian liquid.

\section{ Mode relaxation times in the 4.5 mg/ml HA solution}
By approximating the autocorrelation of mode amplitudes as single exponential s $\langle a_q(t+\delta)\cdot a_q(t)\rangle=|a_q(t)\cdot a_q(t)|\exp(-\delta/\tau_q)$, we can roughly estimate the relaxation time of the $k$th mode as $\tau_q=\gamma_{\mbox{\scriptsize eff}}/\kappa q^4=k_0\eta_{\mbox{\scriptsize eff}}/\kappa q^4$, where the effective viscosity ($\eta_{\mbox{\scriptsize eff}}$) can be estimated by fitting $G''$ with $G''\sim\eta_{\mbox{\scriptsize eff}}\omega$ \cite{2003_Rubinstein}. For the 19 $\mu$m-long SWNT we obtain relaxation times $\tau_q$ for 1st and 2nd modes of 857s and 111s, respectively. These relaxation times are longer than or close to the recording time (154 s). Therefore, these modes don't completely relax during the recording time, and complex shear moduli cannot be calculated from modes 1 to 4 of of this SWNT, and are not shown in Fig. 3(b).

\section{Error analysis of filament microrheology}
\begin{figure}[H]
\centering 
\includegraphics[width=90mm]{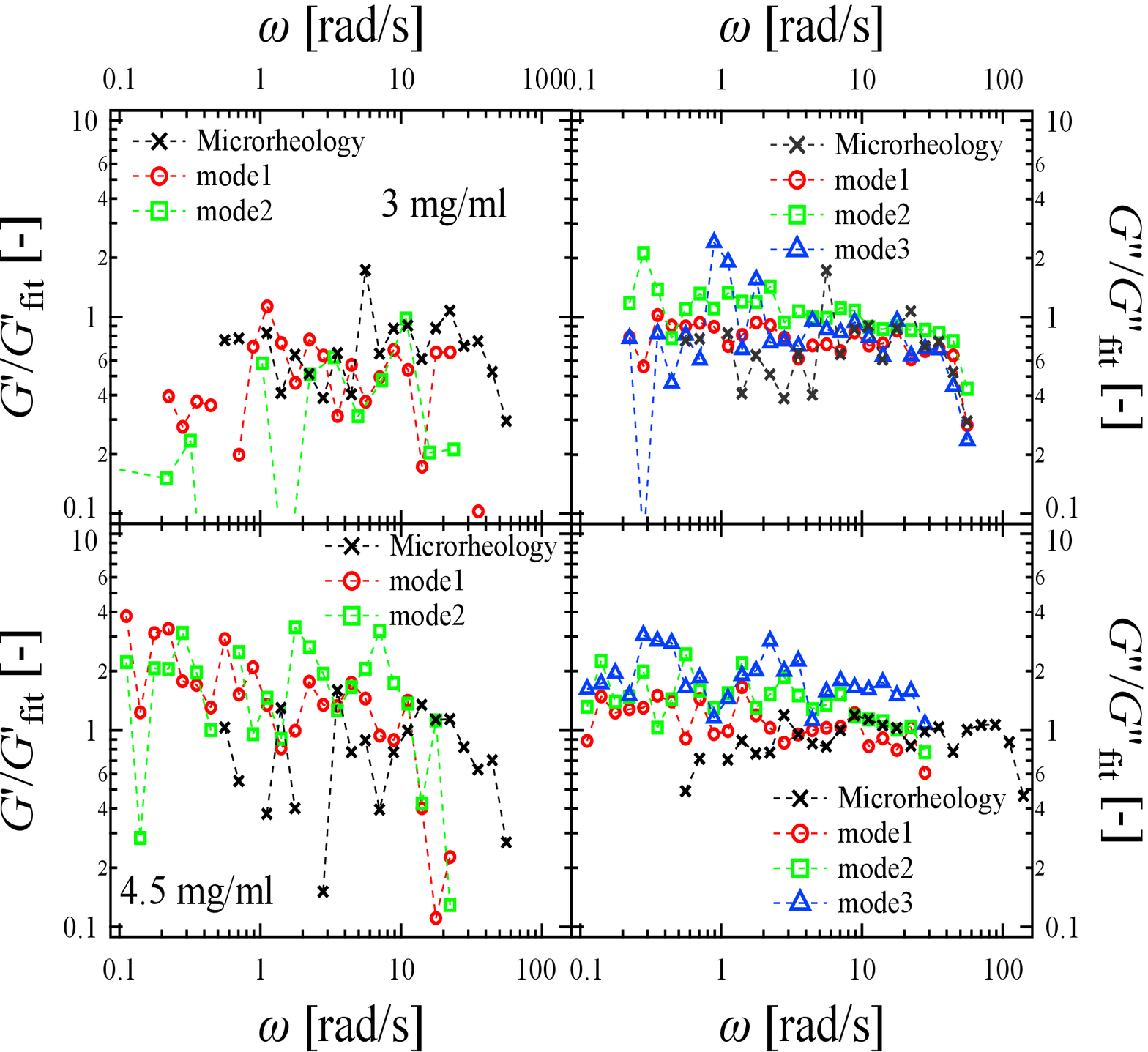}
\caption{Viscoelasticity of 3 and 4.5 mg/ml HA solution measured by FMR, normalized by the power-law fitting results of bead MR results. For this plot, we analyzed 18 recordings from 5 SWNTs with lengths of 4.5 - 8.5 $\mu$m in 3 mg/ml and 14 recordings from 8 SWNTs with lengths of 4.5 - 7.5 $\mu$m in 4.5 mg/ml.}
\label{fig_Gfit}
\end{figure}

To check on the agreement between $G_{\rm{bead}}$ and $G_{\rm{filament}}$, $G_{\rm{bead}}$ was fitted by a power law, and $G_{\rm{filament}}$ was divided by the fitting result of $G_{\rm{bead}}$ in Fig. \ref{fig_Gfit}. Normalized complex shear moduli fluctuate around 1 in all the data sets, suggesting that the complex shear moduli evaluated from different filament bending modes coincide with the bead microrheology results.

\section{Comparison of symmetric method with Kramers-Kronig method to calculate complex shear moduli}

\begin{figure}[H]
\centering 
\includegraphics[width=90mm]{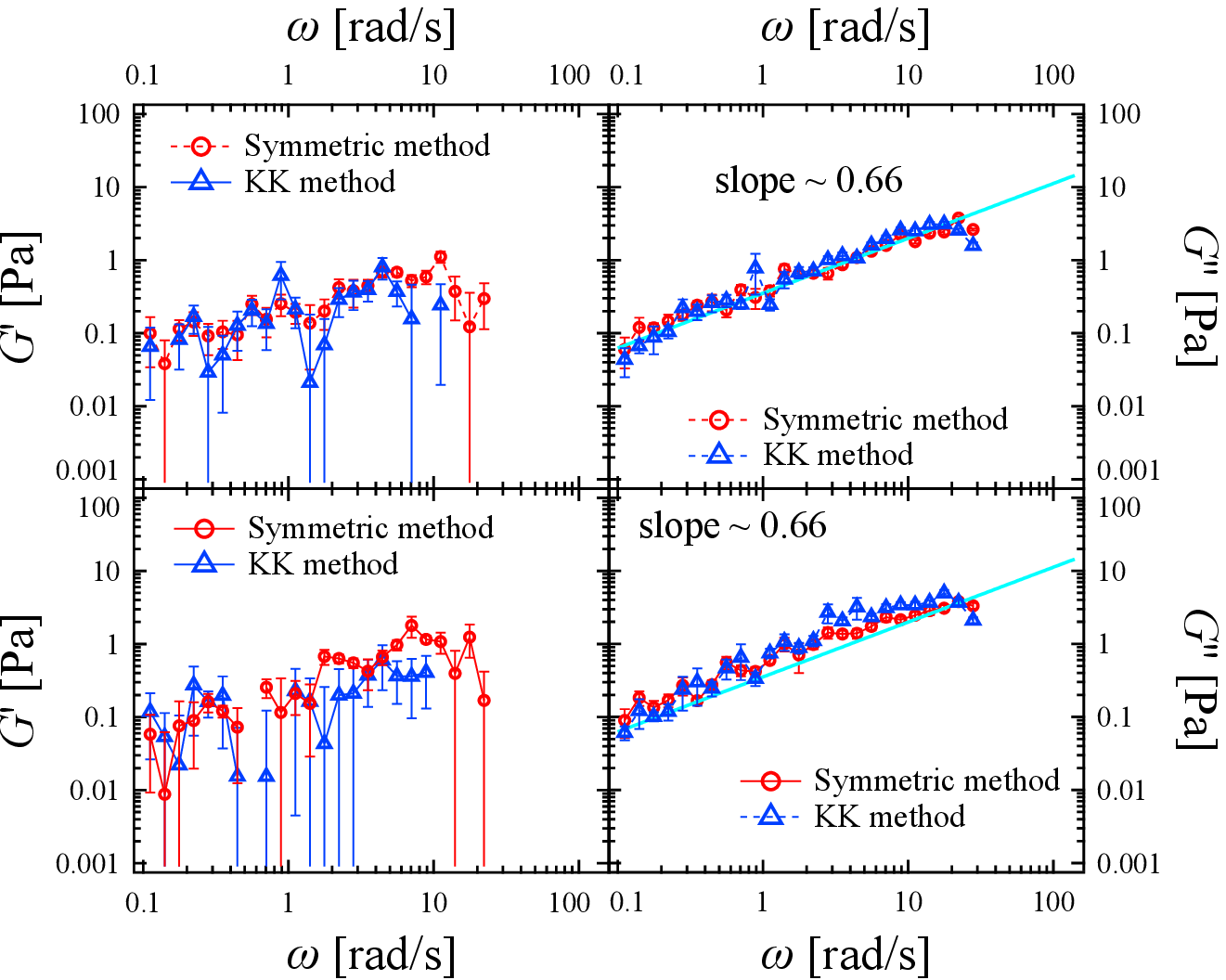}
\caption{Complex shear modulus measured with FMR for 4.5 mg/ml HA solution and evaluated with the KK method and the symmetric method. The bending modes of top and bottom panels are 1st and 2nd modes, respectively. (14 recordings from 8 SWNTs with lengths of 4.5 - 7.5 $\mu$m)}
\label{fig_KKSym}
\end{figure}

\setcounter{equation}{4}
Alternative to the symmetric data evaluation method we used above {\cite{2018_Nishi_SoftMatter}}, $G_{\rm{filament}}$ can also be evaluated using a Kramers-Kronig integral (KK integral) \cite{1997_Gittes_PRL}.
The power spectral density (PSD) of the amplitude fluctuations of the $k$th mode $C_q(\omega)=\langle|a_q(\omega)|^2\rangle$ is directly related to the imaginary part of the response function $\chi_q''(\omega)$ via the FDT \cite{1987_Chandler} as given in Eq. (4).
The real part of the response function $\chi_q'(\omega)$ can then be determined using a KK integral \cite{1997_Gittes_PRL}
\begin{eqnarray}
\chi_q'(\omega)&=&\frac{2}{\pi}\int_0^{\infty}\frac{\xi\chi_q''(\omega)}{\omega^2+\xi^2}d\xi\nonumber\\
&=&\frac{2}{\pi}\int_0^{\infty}\cos(\omega t)dt\int_0^{\infty}\chi_q''(\omega)\sin(\xi t)d\xi
\label{eqKK}
\end{eqnarray}
To speed up this calculation, we performed additional Fourier and inverse Fourier transformations, using fast Fourier transformations (FFT), in the last line.  
We compared the symmetric method with the KK method for the case of the 4.5 mg/ml HA solution in Fig. \ref{fig_KKSym}. 
 The complex shear moduli evaluated from both methods largely superimpose in this case, indicating that the KK method can be used alternatively to the symmetric method.
As pointed out previously \cite{2018_Nishi_SoftMatter}, the main advantage of the symmetric method (Eq. 2) is that it delivers comparable accuracy in both components of $G$ at high-frequencies: high-frequency artefacts primarily showing up in  $G'$ are diminished in the symmetric method based on evaluating $\chi_q(\omega)$. 
Here the symmetric method again provides a small improvement of $G'$ compared to the KK method at high-frequencies, as shown in Fig. \ref{fig_KKSym}, in spite of noisy data sets with a limited number of frames.


\section{Complex shear moduli evaluated from a 19 $\mu$m long SWNT}

\begin{figure}[H]
\centering 
\includegraphics[width=90mm]{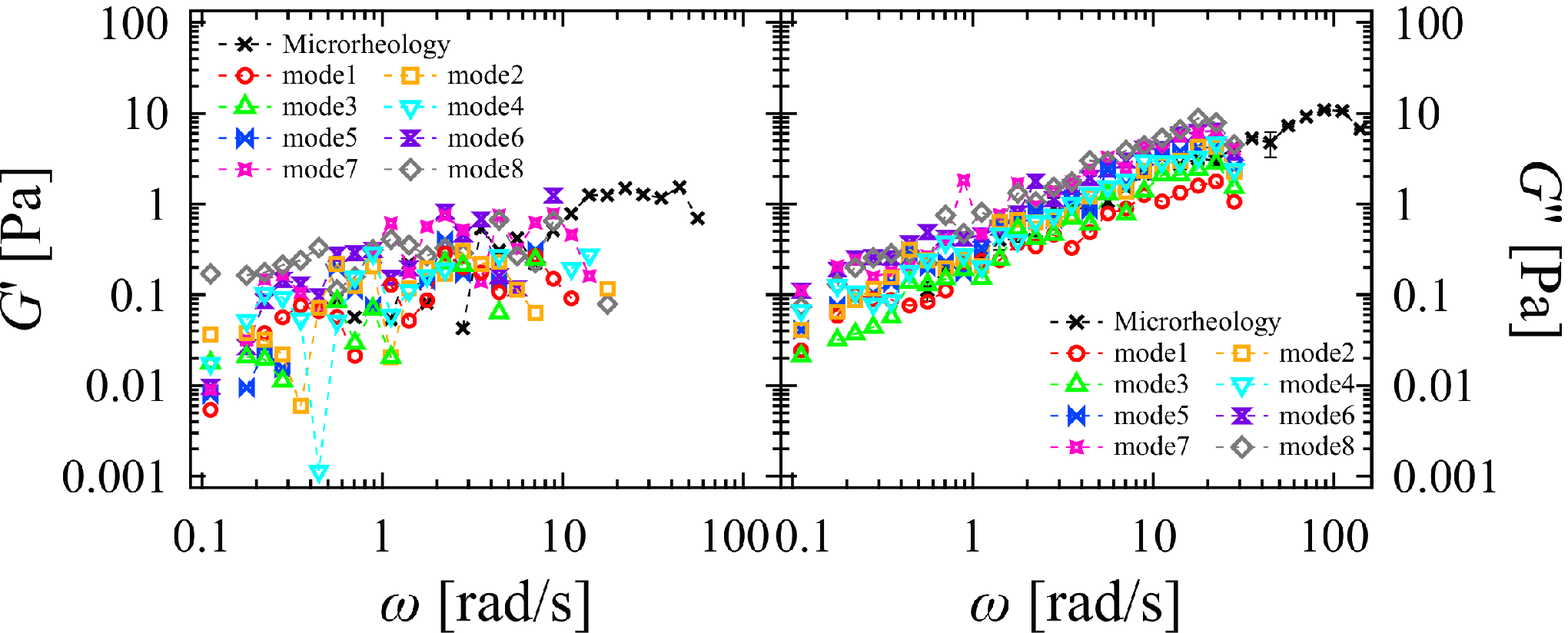}
\caption{Viscoelasticity of 4.5 mg/ml HA solution measured by FMR from the bending dynamics of  a single 19 $\mu$m long SWNT, recorded for 154 s.  The symmetric method was used for data evaluation. }
\label{fig_long}
\end{figure}

An exceptionally long SWNT with a length of 19 $\mu$m was tracked in the 4.5 mg/ml HA solution. The result of mode analysis from the recording of this long SWNT is shown in Fig. \ref{fig_long}.
Although the overall recording time (154 s) was not sufficiently longer than the relaxation time of the first modes (the relaxation time of the 2nd mode was 111 s as discussed above), we plotted all the (apparent) complex shear moduli in Fig. \ref{fig_long} for comparison. As shown in Fig. \ref{fig_long}, the apparent complex shear moduli obtained from modes 1 to 4 are still relatively close to the conventional microrheology data although the recording time was marginal. This result confirms the robustness of the FMR method of simultaneously measuring viscoelasticity over a range of length-scales from one single SWNT.

\end{document}